\documentclass[12pt]{article}

\catcode`\@=11
\@addtoreset{equation}{section}

\global\arraycolsep=2pt

\usepackage{mathrsfs,amsbsy,amssymb,latexsym,amsfonts,amsmath,cite}
\usepackage{graphicx,color}

\newcommand{\al}{\alpha}
\newcommand{\be}{\beta}
\newcommand{\ga}{\gamma}

\newcommand{\ep}{\epsilon}

\newcommand{\ze}{\zeta}

\newcommand{\Tr}{{\rm Tr}}

\newcommand{\alg}[1]{\mathfrak{#1}}
\newcommand{\el}{\nonumber}

\begin{document}

\begin{flushright}
\parbox{4cm}
{KUNS-2489 \\ 
ITP-UU-14/13 \\ 
SPIN-14/13}
\end{flushright}

\vspace*{1.5cm}

\begin{center}
{\Large\bf 
Integrability of classical strings dual for noncommutative gauge theories 
} 
\vspace*{1.5cm}\\
{\large Takuya Matsumoto$^{\dagger}$\footnote{E-mail:~t.matsumoto@uu.nl} 
and Kentaroh Yoshida$^{\ast}$\footnote{E-mail:~kyoshida@gauge.scphys.kyoto-u.ac.jp}} 
\end{center}
\vspace*{0.25cm}
\begin{center}
$^{\dagger}${\it Institute for Theoretical Physics and Spinoza Institute, 
Utrecht University, \\ 
Leuvenlaan 4, 3854 CE Utrecht, The Netherlands.} 
\vspace*{0.25cm}\\ 
$^{\ast}${\it Department of Physics, Kyoto University, \\ 
Kyoto 606-8502, Japan.} 
\end{center}
\vspace{1cm}

\begin{abstract}
We derive the gravity duals of noncommutative gauge theories from the Yang-Baxter 
sigma model description of the AdS$_5\times$S$^5$ superstring with classical $r$-matrices.  
The corresponding classical $r$-matrices are 1) solutions of the classical 
Yang-Baxter equation (CYBE), 2) skew-symmetric, 3) nilpotent and 4) abelian. 
Hence these should be called {\it abelian Jordanian deformations}. 
As a result, the gravity duals are shown to be integrable deformations of AdS$_5\times$S$^5$\,. 
Then, abelian twists of AdS$_5$ are also investigated.  
These results provide a support for the gravity/CYBE correspondence proposed 
in arXiv:1404.1838. 
\end{abstract}

\setcounter{footnote}{0}
\setcounter{page}{0}
\thispagestyle{empty}

\newpage

\section{Introduction}

A particular class of gauge/gravity dualities can be seen as deformations of AdS/CFT \cite{M}. 
With great progress, an integrable structure inhabiting AdS/CFT is well recognized now \cite{review}. 
The Green-Schwarz string action on AdS$_5\times$S$^5$ is constructed from a supercoset \cite{MT} 
\[
PSU(2,2|4)/[SO(1,4)\times SO(5)]
\]
and the classical integrability follows from the $\mathbb{Z}_4$-grading \cite{BPR}\footnote{
For another formulation \cite{RS} of the AdS$_5\times$S$^5$ superstring, 
the classical integrability is argued in \cite{Hatsuda}. For a classification of integrable supercosets, 
see \cite{Zarembo,Wulf}. For an argument on non-symmetric cosets, see \cite{SYY}.}. 
Some deformations of the AdS/CFT correspondence may preserve the integrability and 
hence it would be interesting to consider a method to classify the integrable deformations. 

\medskip 

A possible way is to employ the Yang-Baxter sigma model description \cite{Klimcik,DMV} 
(For $q$-deformed su(2) and its affine extension, see \cite{KYhybrid} and \cite{KMY-QAA}, respectively). 
It has been applied to the AdS$_5\times$S$^5$ superstring in \cite{DMV2}. 
According to this approach,  integrable deformations of AdS$_5\times$S$^5$ are given 
in terms of classical $r$-matrices satisfying modified classical Yang-Baxter equation (mCYBE). 
The case of \cite{DMV2} corresponds to the classical $r$-matrix of Drinfeld-Jimbo type \cite{Drinfeld1,Drinfeld2,Jimbo}. 
The metric in the string frame and NS-NS two-form are obtained \cite{ABF} 
and some generalizations to other cases are discussed in \cite{HRT}. 
It is an intriguing issue to look for the complete gravitational 
solution. A mirror TBA is also proposed \cite{AdLvT}.

\medskip 

As a generalization of the Yang-Baxter sigma model description, one may consider classical Yang-Baxter equation 
(CYBE) rather than mCYBE. The classical action of the AdS$_5\times$S$^5$ superstring has been constructed 
in \cite{KMY-Jordanian-typeIIB}. The integrable deformations are basically regarded as Drinfeld-Reshetikhin twists 
\cite{Drinfeld1,Drinfeld2,R}
including Jordanian twists \cite{Jordanian,KLM} and abelian twists.  
Hence one can classify integrable deformations of this kind 
in terms of classical $r$-matrices. We will refer this picture as to the gravity/CYBE correspondence. 
The first example is presented in \cite{SUGRA-KMY}\footnote{The solution is closely related to the one 
in Appendix C of \cite{HRR}.}. 
As another example, Lunin-Maldacena backgrounds \cite{LM,Frolov} have also been derived \cite{LM-MY}. 

\medskip 

In this note, we derive the gravity duals of noncommutative (NC) gauge theories \cite{HI,MR} 
from the Yang-Baxter sigma model description of the AdS$_5\times$S$^5$ superstring with classical $r$-matrices.  
The corresponding classical $r$-matrices are 1) solutions of CYBE, 
2) skew-symmetric, 3) nilpotent and 4) abelian. Hence these should be called 
{\it abelian Jordanian deformations}. 
As a result, the gravity duals of NC gauge theories  
are shown to be integrable deformations 
of AdS$_5\times$S$^5$\,. Then, abelian twists of AdS$_5$ are also investigated. 
These results provide a support 
for the gravity/CYBE correspondence proposed in \cite{LM-MY}. 

\medskip

This note is organized as follows. 
Section 2 gives a short summary of the Yang-Baxter sigma model description 
of the AdS$_5\times$S$^5$ superstring with classical $r$-matrices satisfying CYBE. 
Then we introduce three classes of skew-symmetric solutions of CYBE. 
A new class of $r$-matrices induces abelian Jordanian deformations. 
Section 3 presents examples of abelian Jordanian type, which lead to the gravity duals of NC gauge theories. 
In section 4, we consider a deformation of AdS$_5$ with an abelian $r$-matrix concerned 
with a TsT transformation of AdS$_5$\,. Section 5 is devoted to conclusion and discussion. 
We argue some implications of this result and future directions in studies of the gravity/CYBE correspondence. 
In Appendix A our notation and convention are summarized. Appendix B presents the gravity duals of NC gauge theories 
with six deformation parameters.  
Appendix C describes the detailed computation of three-parameter abelian twists of AdS$_5$\,.

\section{Integrable deformations of the AdS$_5\times$S$^5$ superstring} 

We introduce here integrable deformations of the AdS$_5\times$S$^5$ superstring 
based on the Yang-Baxter sigma model description with CYBE \cite{KMY-Jordanian-typeIIB}\,. 
After giving a short review on the general form of deformed actions, 
we present three classes of classical $r$-matrices.

\subsection{Deforming the AdS$_5\times$S$^5$ superstring action with CYBE \label{2.1}}

A class of integrable deformations of the AdS$_5\times$S$^5$ superstring 
can be described with classical $r$-matrices satisfying CYBE \cite{KMY-Jordanian-typeIIB}. 
The deformed action is given by  
\begin{eqnarray}
S=-\frac{1}{4}(\ga^{\al\be}-\ep^{\al\be})\int^\infty_{-\infty}\!\!\!d\tau\int^{2\pi}_0\!\!\!d\sigma~
{\rm Str}\Bigl(A_\al d\circ\frac{1}{1-\eta R_g\circ d}A_\be\Bigr)\,,  
\label{action}
\end{eqnarray}
where the left-invariant one-form $A_\alpha$ is defined as 
\begin{eqnarray}
A_\alpha\equiv g^{-1}\partial_\alpha g\,, \qquad 
g\in SU(2,2|4)\,. 
\end{eqnarray}
Here $\ga^{\al\be}$ and $\ep^{\al\be}$ are the flat metric and the 
anti-symmetric tensor on the string world-sheet.  
The operator $R_g$ is defined as 
\begin{align}
R_g(X)\equiv g^{-1}R(gXg^{-1})g\,,  
\end{align} 
where a linear operator $R$ satisfies CYBE rather than mCYBE \cite{DMV2}\,. 
The R-operator is related to the tensorial representation of classical $r$-matrix through 
\begin{align}
&R(X)=\Tr_2[r(1\otimes X)]=\sum_i \bigl(a_i\Tr(b_iX)-b_i\Tr(a_iX)\bigr) 
\label{linearR} \\
&\text{with}\quad r=\sum_i a_i\wedge b_i\equiv \sum_i (a_i\otimes b_i-b_i\otimes a_i)\,. \el 
\end{align}
The operator $d$ is given by the following,  
\begin{align}
d=P_1+2P_2-P_3\,,  
\label{d}
\end{align}
where $P_i$ ($i=0,1,2,3$) are the projections to the $\mathbb{Z}_4$-graded 
components of $\alg{su}(2,2|4)$\,.  
$P_0\,,P_2$ and $P_1\,,P_3$ are the projectors to the bosonic and 
fermionic generators, respectively. 
In particular, $P_0(\alg{su}(2,2|4))$ is nothing but $\alg{so}(1,4)\oplus\alg{so}(5)$\,. 

\medskip 

For the action \eqref{action} with an R-operator satisfying CYBE, 
the Lax pair has been constructed \cite{KMY-Jordanian-typeIIB} 
and the classical integrability is ensured in this sense. 
The $\kappa$-invariance has been proven as well \cite{KMY-Jordanian-typeIIB}\,. 

\subsection{A classification of classical $r$-matrices} 

According to the construction of the deformed string action, 
one may expect the correspondence between integrable deformations of AdS$_5\times$S$^5$ 
and classical $r$-matrices, called the gravity/CYBE correspondence \cite{LM-MY}\,. 
To study along this direction, it would be valuable to classify some 
typical class of skew-symmetric solutions of CYBE. 

\medskip 

There are three types of classical $r$-matrices:  
i) Jordanian, ii) abelian, and iii) abelian Jordanian. 
In particular,  the third class will play a crucial role in the next section.  

\medskip 

In order to study deformations of AdS$_5$ later, 
let us consider the case of $\alg{su}(2,2)$\,.   

\medskip 

\subsubsection*{i) Jordanian $r$-matrix.}  

The first class is classical $r$-matrices of Jordanian type, 
\begin{align}
r_{\rm Jor}=E_{ij}\wedge (E_{ii}-E_{jj}) -2 \sum_{i<k<j}E_{ik}\wedge E_{kj}
\qquad (1\leq i<j\leq4) \,, 
\label{Jor}
\end{align}
where $(E_{ij})_{kl} \equiv \delta_{ik}\delta_{jl}$ 
are the fundamental representation of $\alg{su}(2,2)$\,. 
The characteristic property of Jordanian type $r$-matrices is the nilpotency.
Indeed, we could verify that the associated linear R-operator exhibits 
$\left(R_{\rm Jor}\right)^n=0$ for $n\geq3$\,.  

\medskip 

Jordanian deformations of the AdS$_5\times$S$^5$ superstring are considered in \cite{KMY-Jordanian-typeIIB}\,. 
A simple example of the corresponding type IIB supergravity solution is presented in \cite{SUGRA-KMY}\,. 
Only the AdS$_5$ part is deformed and it contains a three-dimensional Schr\"odinger spacetime as a subspace. 
Hence it may be regarded as a generalization of \cite{KY-Sch,Jordanian-KMY,Kame}. 
It seems likely that the resulting metric is closely related to a null Melvin twist \cite{HRR}.

\subsubsection*{ii) Abelian $r$-matrix.}  

The second class is  abelian $r$-matrices composed of the Cartan generators as follows:   
\begin{align}
r_{\rm Abe}=\sum_{1\leq i <j\leq3}\mu_{ij}\,(E_{ii}-E_{i+1,i+1})\wedge(E_{jj}-E_{j+1,j+1})\,,  
\end{align}
where $\mu_{ij}=-\mu_{ji}$ are arbitrary parameters. 
Since these commute with each other and hence satisfy CYBE obviously. 
The abelian $r$-matrix is a particular example of the Drinfeld-Reshetikhin 
twists \cite{Drinfeld1,Drinfeld2,R}. 
Note that abelian $r$-matrices are intrinsic to higher rank cases (rank $\geq 2$). 

\medskip 

It has been shown in \cite{LM-MY} that abelian $r$-matrices lead to 
$\ga$-deformed backgrounds \cite{Frolov}, which include   
the Lunin-Maldacena background \cite{LM} as a particular case. 
In section 4, we will consider an abelian twist of AdS$_5$ with a single parameter. 
For multi-parameter cases, see Appendix C\,.

\subsubsection*{iii) Abelian Jordanian $r$-matrix.}  

The third class is composed of $r$-matrices which are nilpotent {\it and} abelian. 
These should be called {\it abelian Jordanian} $r$-matrices. 
A typical example takes the following form,  
\begin{align}
r_{\rm AJ}=\sum_{\substack{i,k=1,2\\ j,l = 3,4}}\nu_{(ij),(kl)}\, E_{ij}\wedge E_{kl}\,,   
\end{align}
with arbitrary parameters $\nu_{(ij),(kl)}=-\nu_{(kl),(ij)}$\,. 
Because $E_{ij}$ ($i=1,2\,, j= 3,4$) are the positive root generators
and commute with each other, the square of the associated R-operator 
already vanishes like, 
\[
\left(R_{\rm AJ} \right)^2=0\,, 
\]
in comparison to Jordanian $r$-matrices \eqref{Jor} 
for which $\left(R_{\rm Jor}\right)^2\neq 0$ and $\left(R_{\rm Jor}\right)^3=0$ 
in general. 

\medskip 

In the next section, we will show that classical $r$-matrices of abelian Jordanian type 
correspond to the gravity duals of NC gauge theories \cite{HI,MR}.

\section{Examples - gravity duals of NC gauge theories}

Let us consider examples of classical $r$-matrices of abelian Jordanian type. 
These lead to the gravity duals of NC gauge theories \cite{HI,MR}\,. 
Hereafter we will concentrate on the AdS$_5$ part and  S$^5$ is not deformed. 

\medskip 
  
A possible example is given by 
\begin{align}
r_{\rm AJ}=\mu \, p_2\wedge p_3+\nu \, p_0\wedge p_1\,, \label{r-2para}
\end{align}
where $\mu,\nu$ are deformation parameters. Here $p_\mu$ ($\mu=0,1,2,3$) 
are the upper triangular matrices defined as 
\begin{align}
p_\mu \equiv \frac{1}{2}\ga_\mu-m_{\mu5}\,. 
\label{pmu}
\end{align}
For our convention of $\ga_\mu$ and the $\alg{su}(2,2)$ generators\,, see Appendix A. 
It should be emphasized that
$p_\mu$'s are upper triangular and satisfy the following property: 
\begin{align}
p_\mu p_\nu=0\,.  
\end{align}
Thus the classical $r$-matrix \eqref{r-2para} 
is of abelian Jordanian type and 
trivially satisfies CYBE. 

\medskip 

To evaluate the Lagrangian \eqref{action}\,, 
let us take the following coset parametrization \cite{SUGRA-KMY}\,:   
\begin{align}
g_{\rm s}=\exp\left[p_0\,x^0+p_1\,x^1+p_2\,x^2+p_3\,x^3\right]\exp\left[\frac{\ga_5}{2}\log z\right]
\quad \in~ SU(2,2)/SO(1,4)\,.  
\end{align}
Then the AdS$_5$ part of \eqref{action} can be rewritten  as 
\begin{align}
&L =-\frac{1}{2} (\ga^{\al\be}-\ep^{\al\be}){\rm Tr}
\left[A_\alpha P_2(J_\be) \right] 
\label{action-ads} \\
&\text{with}\qquad 
J_\be\equiv \frac{1}{1-2\eta\bigl[R_{\rm AJ}\bigr]_g\circ P_2}A_\be\,. 
\end{align}
Here $A_\al=g^{-1}\partial_\al g$ is restricted to $\alg{su}(2,2)$ and 
the associated R-operator $R_{\rm AJ}$ with \eqref{r-2para} 
is determined by the relation \eqref{linearR}\,. 

\medskip 

It is convenient to divide the Lagrangian $L$ into two parts like $L=L_G+L_B$\,, 
where $L_G$ is the metric part and $L_B$ is the coupling to an NS-NS two-form, respectively:  
\begin{eqnarray}
L_G &\equiv& \frac{1}{2} [\Tr(A_{\tau}P_2(J_{\tau}))-\Tr(A_{\sigma}P_2(J_{\sigma}))] \,, \nonumber \\ 
L_B &\equiv& \frac{1}{2} [\Tr(A_{\tau}P_2(J_{\sigma}))-\Tr(A_{\sigma}P_2(J_{\tau}))] \,.  
\label{LGLB}
\end{eqnarray}

\medskip 

To derive the explicit form of $L$\,,  
it is sufficient to compute the projected current $P_2(J_\al)$ 
rather than $J_\al$ itself.  
Hence the computation is reduced to solving the following equation,  
\begin{align}
\left(1-2\eta P_2\circ \left[R_{\rm AJ}\right]_g\right)P_2(J_\al)=P_2(A_\al)\,. 
\label{equation}
\end{align}
Note that $P_2(A_\alpha)$ is expanded with $\gamma$ matrices as follows:    
\begin{align}
P_2(A_\al)&=\frac{\partial_\al x^0\,\ga_0+\partial_\al x^1\,\ga_1
+\partial_\al x^2\,\ga_2+\partial_\al x^3\,\ga_3+\partial_\al z\,\ga_5}{2z}\,. 
\label{P_2(A)} 
\end{align}
Then, by combining \eqref{P_2(A)} with \eqref{equation}\,, $P_2(J_\alpha)$ can be obtained as 
\begin{align}
P_2(J_\alpha)
&=\frac{z(z^2\partial_\al x_0+2\eta\nu \partial_\al x_1)}{2(z^4-4\eta^2\nu^2)}\ga_0
+ \frac{z( z^2 \partial_\al x_1+2\eta \nu \partial_\al x_0 )}{2(z^4-4\eta^2\nu^2)}\ga_1
\el\\
&\quad 
+ \frac{z(z^2\partial_\al x_2 + 2  \eta\mu \partial_\al x_3)}{2(z^4+4\eta^2\mu^2)}\ga_2 
+ \frac{z(z^2\partial_\al x_3-2\eta\mu\partial_\al x_2)}{2(z^4+4\eta^2\mu^2)}\ga_3 
+ \frac{\partial_\al z}{2z}\ga_5\,. 
\end{align}

\medskip 

The resulting forms of $L_G$ and $L_B$ are given by, respectively,   
\begin{align}
L_G &= -\frac{\ga^{\al\be}}{2} \Bigl[
\frac{z^2 (-\partial_\al x_0\partial_\be x_0 + \partial_\al x_1\partial_\be x_1)}
{z^4-4\eta^2 \nu^2}
+\frac{z^2 (\partial_\al x_2\partial_\be x_2 + \partial_\al x_3\partial_\be x_3)}
{z^4+4\eta^2 \mu^2}
+\frac{\partial_\al z\partial_\be z}{z^2}\Bigr]\,,  
\label{LG3}
\\
L_B&=\ep^{\al\be} \Bigl[
-\frac{2\eta\nu}{z^4 - 4\eta^2\nu^2}\partial_\al x_0\partial_\be x_1 
+\frac{2\eta\mu}{z^4 + 4\eta^2\mu^2}\partial_\al x_2\partial_\be x_3 \Bigr]\,.
\label{LB3}
\end{align}
Here two deformation parameters $\mu,\nu$ and 
one normalization factor $\eta$ are contained. 

\medskip 

It is easy to see the metric and the NS-NS two-form from \eqref{LG3} and \eqref{LB3}\,.
By introducing new parameter $a$ and $a'$ through the identification, 
\begin{align} 
2\eta\,\mu =a^2\,,  \qquad 2\eta\,\nu =i a'^2 \,, 
\end{align}
one can find that the resulting metric and two-form exactly agree 
with the ones of the gravity duals of NC gauge theories presented in \cite{HI,MR},  
up to the coordinate change $z=1/u$ and the Wick rotation $x_0\to ix_0$\,. 
This result shows that the gravity duals of NC gauge theories \cite{HI,MR} are 
integrable deformation of AdS$_5$\,.

\section{Abelian twists of AdS$_5$} 

As another kind of integrable deformation of AdS$_5$\,, 
we consider an abelian twist of AdS$_5$ with a single parameter\footnote{
Abelian twists of S$^5$ have been studied in \cite{LM-MY}. 
The resulting geometries are three-parameter $\gamma$-deformed S$^5$ \cite{LM,Frolov}. 
}. 
For a three-parameter generalization, see Appendix C\,. 

\medskip 

Let us consider an abelian $r$-matrix, 
\begin{eqnarray}
r_{\rm Abe}^{(\mu)} = \mu\, h_1 \wedge h_2 \,,
\label{rabe-1}
\end{eqnarray}
with a deformation parameter $\mu$\,. Here  
$h_i~(i=1,2)$ are two of the Cartan generators of $\mathfrak{su}(2,2)$ 
and belong to the fundamental representation,  
\begin{align}
h_1={\rm diag}(-1,1,-1,1)\,,\quad 
h_2={\rm diag}(-1,1,1-,1)\,.  
\end{align}
Then, the AdS$_5$ part of the Lagrangian \eqref{action} is given by  
\begin{align}
&L=L_G+L_B=
-\frac{1}{2} (\ga^{\al\be}-\ep^{\al\be}){\rm Tr}\left[A_\al P_2(J_\be)\right]\,, \\
\text{with} \quad &
J_\be\equiv \frac{1}{1-2\eta\bigl[R_{\rm Abe}^{(\mu)}\bigr]_g\circ P_2}A_\be\,, 
\end{align}
where the current $A_\al$ is $\alg{su}(2,2)$-valued and the R-operator associated 
with \eqref{rabe-1} is defined by the rule \eqref{linearR}. 

\medskip 

The projected current $P_2(J_\al)$ is to be determined by solving the equation,  
\begin{align}
\left(1-2\eta P_2\circ \left[R_{\rm Abe}^{(\mu)}\right]_g\right)P_2(J_\al)=P_2(A_\al)\,. 
\label{eqAJ}
\end{align}
By using the coset parameterization (\ref{coset-para})\,,  
$P_2(A_\alpha)$ is expanded with respect to $\gamma$ matrices,     
\begin{align}
P_2(A_\al)&=\frac{1}{2}\Bigl[
-\partial_\al \rho \,\ga_1
+ i \cosh\rho\, \partial_\al\psi_3\, \ga_5\el\\
&\qquad 
-\sinh\rho\,( \cos\ze\,  \partial_\al\psi_1 \,\ga_2
+  \partial_\al\ze \,\ga_3
- i  \sin\ze\, \partial_\al \psi_2 \, \ga_0) \Bigr]\,. 
\label{P2A} 
\end{align}
Then, by plugging \eqref{P2A} with \eqref{eqAJ}\,, 
$P_2(J_\alpha)$ can be obtained as 
\begin{align}
P_2(J_\alpha)=j_\alpha^0\,\ga_0+j_\alpha^1\,\ga_1+j_\alpha^2\,\ga_2 
+j_\alpha^3\,\ga_3 +j_\alpha^5\,\ga_5\,, 
\end{align}
with the coefficients 
\begin{align}
&j_\alpha^0=\frac{i}{2}
\frac{ \sin\ze \sinh\rho}{1 + 16 \eta^2\mu^2 \sin^22\ze \sinh^4\rho} 
\left(
\partial_\al\psi_2  
+8\eta\mu\cos^2\ze\sinh^2\rho\partial_\al\psi_1\right)\,,  
\el \\ 
&j_\alpha^1=-\frac{1}{2}\partial_\al \rho\,, 
\el \\
&j_\alpha^2=-\frac{1}{2}
\frac{\cos\ze\sinh\rho}{1 + 16 \eta^2\mu^2 \sin^22\ze \sinh^4\rho} 
\left(\partial_\al \psi_1 - 8 \eta\mu \sin^2\ze \sinh^2\rho \partial_\al\psi_2
\right)
\,, \nonumber \\
&j_\alpha^3=-\frac{1}{2}\sinh\rho\, \partial_\al\ze\,, \el\\
&j_\alpha^5=\frac{i}{2} \cosh \rho\partial_\al\psi_3\,. 
\end{align}
Finally, the resulting expressions of $L_G$ and $L_B$ are given by, respectively,   
\begin{align}
L_G &= -\frac{\ga^{\al\be}}{2} \Bigl[
\sinh^2\rho \partial_\al\ze\partial_\be\ze 
+ \partial_\al\rho\partial_\be\rho  
- \cosh^2\rho \partial_\al\psi_3\partial_\be\psi_3 
\el\\ 
&\qquad \qquad 
+\frac{ \sinh^2\rho}{1 + \hat\ga^2 \sin^2\ze\cos^2\ze \sinh^4\rho}
(\cos^2\ze \partial_\al\psi_1\partial_\be\psi_1
+ \sin^2\ze \partial_\al\psi_2\partial_\be\psi_2)
\Bigr]\,, \label{LG3abe1}
\\
L_B&=-\ep^{\al\be}\frac{\hat\ga \cos^2\ze \sin^2\ze \sinh^4\rho}
{ 1 + \hat\ga^2\cos^2\ze \sin^2\ze \sinh^4\rho}
 \partial_\al\psi_1 \partial_\be\psi_2\,. 
\label{LB3abe1}
\end{align}
Here a new deformation parameter $\hat\ga$ is defined as 
\begin{align} 
\hat\ga\equiv 8\eta\,\mu \,. 
\end{align}

\medskip 

Now one can read off the metric and NS-NS two-form 
from \eqref{LG3abe1} and \eqref{LB3abe1}\,. By performing the coordinate transformation,    
\begin{align}
\rho_1=\cos\ze \sinh\rho\,, \qquad 
\rho_2=\sin\ze \sinh\rho \,, \qquad
\rho_3=i \cosh\rho\,,     
\label{coord-trf}
\end{align}
the resulting metric and NS-NS two-form are given by  
\begin{eqnarray}
ds^2 &=&d\rho_1^2+d\rho_2^2+d\rho_3^2
+\frac{\rho_1^2d\psi_1^2+\rho_2^2d\psi_2^2}{1 + \hat{\ga}^2\,\rho_1^2\rho_2^2 } 
+\rho_3^2d\psi_3^2
+ds^2_{\rm S_5}\,,   
\label{1-metric} 
\\ 
B_2 &=& \frac{\hat{\ga}\,\rho_1^2\rho_2^2}{ 1 + \hat{\ga}^2\,\rho_1^2\rho_2^2}
d\psi_1\wedge d\psi_2\,.  
\label{1-NSNS}
\end{eqnarray}
Here there is a constraint $\sum_{i=1}^3 \rho_i^2=-1$\,.  

\medskip 

These expressions are quite similar to a one-parameter 
$\gamma$-deformed S$^5$ \cite{Frolov,LM} and thus the solution with the metric (\ref{1-metric}) and the NS-NS two-form (\ref{1-NSNS})  
may be regarded as a single parameter $\ga$-deformation of AdS$_5$\,.

\section{Conclusion and discussion}

We have shown that the gravity duals of NC gauge theories \cite{HI,MR} can be derived 
from the Yang-Baxter sigma model description of the AdS$_5\times$S$^5$ superstring with classical $r$-matrices.  
The corresponding classical $r$-matrices are 1) solutions of CYBE, 2) skew-symmetric, 3) nilpotent and 4) abelian. 
These should be called {\it abelian Jordanian deformations}. 
As a result, the gravity duals are found to be integrable deformations of AdS$_5\times$S$^5$\,. 
Then, abelian twists of AdS$_5$ have also been investigated.  
These results provide a support for the gravity/CYBE correspondence proposed 
in \cite{LM-MY}. 

\medskip 

Our main result here is the integrability of $\mathcal{N}$=4 super Yang-Mills (SYM) theory on noncommutative (NC) spaces. 
Now there are an enormous amount of arguments on the integrability for scattering amplitudes 
of $\mathcal{N}$=4 SYM. 
Integrable deformations of it would be found on NC spaces. 
Our analysis has revealed a relation between classical $r$-matrices and 
deformations parameters of NC spaces. There may be a close connection 
to deformation quantization of Kontsevich \cite{K}. 
Thus one may expect a deep mathematical structure 
behind the correspondence. 
We hope that our result could shed light on new fundamental aspects of integrable deformations.

\subsection*{Acknowledgments}

We would like to thank Io Kawaguchi for useful discussions and collaborations at the earlier stage. 
T.M.\ also thanks Gleb Arutyunov and Riccardo Borsato for useful discussions.  
T.M.\ is supported by the Netherlands Organization for Scientific 
Research (NWO) under the VICI grant 680-47-602.  
T.M.'s work is also part of the ERC Advanced grant research programme 
No.~246974, ``Supersymmetry: a window to non-perturbative physics" 
and of the D-ITP consortium, a program of the NWO that is funded by the 
Dutch Ministry of Education, Culture and Science (OCW).

\appendix

\section*{Appendix} 

\section{Notation and convention}

We shall here summarize our notation and convention, which basically follow \cite{AF-review}\,. 

\medskip 

An element of $\mathfrak{su}(2,2|4)$ is identified with 
an $8\times 8$ supermatrix, 
\begin{eqnarray}
M=\begin{bmatrix}
~m~&~\xi~\\
~\zeta~&~n~
\end{bmatrix}\,. 
\end{eqnarray}
Here $m$ and $n$ are $4\times 4$ matrices with Grassmann even elements, 
while $\xi$ and $\zeta$ are $4\times 4$ matrices with Grassmann odd elements. 
These matrices satisfy a reality condition. Then $m$ and $n$ belong to $\mathfrak{su}(2,2)=\mathfrak{so}(2,4)$ and  
$\mathfrak{su}(4)=\mathfrak{so}(6)$\,, respectively.  

\medskip 

We are concerned with deformations of AdS$_5$\,.  
An explicit basis of $\mathfrak{su}(2,2)$ is the following. 
The $\gamma$ matrices are given by 
\begin{eqnarray}
&&\gamma_1=
\begin{bmatrix}
~0~&~0~&~0~&-1\\
~0~&~0~&~1~&~0\\
~0~&~1~&~0~&~0\\
-1~&~0~&~0~&~0\\
\end{bmatrix}\,, \quad
\gamma_2=
\begin{bmatrix}
~0~&~0~&~0~&~i~\\
~0~&~0~&~i~&~0~\\
~0~&-i~&~0~&~0~\\
-i~&~0~&~0~&~0~\\
\end{bmatrix}\,, \quad 
\gamma_3=
\begin{bmatrix}
~0~&~0~&~1~&~0~\\
~0~&~0~&~0~&~1~\\
~1~&~0~&~0~&~0~\\
~0~&~1~&~0~&~0~\\
\end{bmatrix}\,, \nonumber \\
&&\gamma_0=
\begin{bmatrix}
~0~&~0~&1~&~0~\\
~0~&~0~&~0~&-1~\\
-1~&~0~&~0~&~0~\\
~0~&~1~&~0~&~0~\\
\end{bmatrix}\,, \quad
\gamma_5=i\gamma_1\gamma_2\gamma_3\gamma_0=
\begin{bmatrix}
~1~&~0~&~0~&~0~\\
~0~&~1~&~0~&~0~\\
~0~&~0~&-1~&~0~\\
~0~&~0~&~0~&-1~\\
\end{bmatrix}
\end{eqnarray}
and satisfy the Clifford algebra 
\begin{eqnarray}
\left\{\ga_\mu,\ga_\nu\right\}=2\eta_{\mu\nu}\,,\qquad 
\left\{\ga_\mu,\ga_5\right\}=0 \,, \qquad 
(\ga_5)^2=1\,.   
\end{eqnarray}
The Lie algebra $\alg{so}(1,4)$ is formed by the generators 
\begin{align}
m_{\mu\nu}=\frac{1}{4}\left[\ga_\mu,\ga_\nu\right]\,, \qquad 
m_{\mu5}=\frac{1}{4}\left[\ga_\mu,\ga_5\right] \qquad 
(\mu,\nu=0,1,2,3\,)\,, 
\end{align}
and then $\alg{so}(2,4)=\alg{su}(2,2)$ is spanned by the following set: 
\begin{align}
m_{\mu\nu}\,, \quad m_{\mu5}\,, \quad \ga_\mu\,, \quad \ga_5\,. 
\end{align}

\section{Multi-parameter deformations of AdS$_5$} 

We present here multi-parameter deformations of AdS$_5$ by using 
the Yang-Baxter sigma model description with classical $r$-matrices. 
These may be regarded as a multi-parameter generalization of the gravity duals of NC gauge 
theories discussed in \cite{HI,MR}. 
In the original construction \cite{HI,MR} based on twisted T-dualities, 
it would be intricate to perform T-dualities many times. 
A technical advantage of the Yang-Baxter sigma model description is that 
a single $r$-matrix gives the corresponding metric and NS-NS two-form in a more direct way. 

\medskip 

Let us consider the following classical $r$-matrix, 
\begin{align}
r_{\rm AJ}
&=\mu_1\,p_2\wedge p_3 + \mu_2\,p_3\wedge p_1 + \mu_3\,p_1\wedge p_2  \el \\
&\quad 
+ \nu_1\,p_0\wedge p_1 + \nu_2\,p_0\wedge p_2+ \nu_3\,p_0\wedge p_3\,,   
\end{align}
where $\mu_1,\mu_2,\mu_3$ and $\nu_1,\nu_2,\nu_3$ are six deformation parameters, 
and $p_\mu$ are defined in \eqref{pmu}\,. By following the analysis in section 3\,, 
it is straightforward to get the deformed string action. For simplicity, 
we shall write down only the resulting metric and NS-NS two-form,   
\begin{align}
ds^2 &= 
\frac{dz^2}{z^2}+z^2G\Bigl[
-(z^4+4\eta^2(\mu_1^2+\mu_2^2+\mu_3^2))dx_0^2
+(z^4+4\eta^2(\mu_1^2-\nu_2^2-\nu_3^2))dx_1^2 \el \\
&\qquad \qquad \qquad
+(z^4+4\eta^2(\mu_2^2-\nu_3^2-\nu_1^2))dx_2^2
+(z^4+4\eta^2(\mu_3^2-\nu_1^2-\nu_2^2))dx_3^2 \el \\
&\quad 
-8\eta^2 \bigl[(\mu_2\nu_3-\mu_3\nu_2)dx_0dx_1
+(\mu_3\nu_1-\mu_1\nu_3)dx_0dx_2
+(\mu_1\nu_2-\mu_2\nu_1)dx_0dx_3 \el \\
& \qquad \quad 
-(\mu_1\mu_2+\nu_1\nu_2)dx_1dx_2
-(\mu_2\mu_3+\nu_2\nu_3)dx_2dx_3
-(\mu_1\mu_3+\nu_1\nu_3)dx_1dx_3 \bigr] \Bigr]\,,
\label{LG6}
\\
B_2&=2\eta\, G \bigl[
(z^4 \mu_1-\eta^2\nu_1K) dx_2\wedge dx_3 
-(z^4 \nu_1+\eta^2\mu_1K) dx_0\wedge dx_1
\el\\
&\qquad\quad 
+(z^4 \mu_2-\eta^2\nu_2K) dx_3\wedge dx_1
-(z^4 \nu_2+\eta^2\mu_2K) dx_0\wedge dx_2 
\el\\
&\qquad\quad 
+(z^4 \mu_3-\eta^2\nu_3K) dx_1\wedge dx_2 
-(z^4 \nu_3+\eta^2\mu_3K) dx_0\wedge dx_3
\bigr]\,. 
\label{LB6} 
\end{align}
Here a scalar function $G$ and a constant parameter $K$ are defined as 
\begin{align}
G^{-1}& \equiv 
z^8+4\eta^2z^4(\mu_1^2+\mu_2^2+\mu_3^2-\nu_1^2-\nu_2^2-\nu_3^2)-\eta^4 K^2 
\,, \el \\
K& \equiv 4(\mu_1\nu_1+\mu_2\nu_2+\mu_3\nu_3)\,. \el 
\end{align}
By taking the following identification of the parameters   
\begin{align} 
2\eta\,\mu_1 =a^2\,,  \quad 2\eta\,\nu_1 =i a'^2 \,, 
\quad 
\mu_2=\mu_3=\nu_2=\nu_3=0\,, 
\end{align}
and performing a Wick rotation $x_0\to ix_0$\,, one can reproduce the metric and 
NS-NS two-form of the two-parameter case \cite{HI,MR}. 

\medskip 

Note that the metric \eqref{LG6} and NS-NS two-form \eqref{LB6} 
are complemented with the other field components and gives a complete solution of type IIB supergravity. 
It gives a consistent string background because it is basically obtained by performing a chain of (twisted) 
T-dualities for AdS$_5$\,.

\section{Three-parameter abelian twists of AdS$_5$}

Let us consider here a three-parameter generalization of 
the abelian deformation of AdS$_5$ discussed in section 4\,. 

\medskip 

We will consider the following classical $r$-matrix, 
\begin{eqnarray}
r_{\rm Abe}^{(\mu_1,\mu_2,\mu_3)} = \mu_3\, h_1 \wedge h_2 
+ \mu_1\, h_2 \wedge h_3 + \mu_2\, h_3 \wedge h_1\,,
\label{3-para}
\end{eqnarray}
with deformation parameters $\mu_i$\,. 
Here $h_i$ are  the three 
Cartan generators of $\mathfrak{su}(2,2)$ and belong to the fundamental representation, 
\begin{align}
h_1={\rm diag}(-1,1,-1,1)\,,\quad 
h_2={\rm diag}(-1,1,1-,1)\,,\quad 
h_3={\rm diag}(1,1,-1,-1)\,. 
\end{align}

\medskip 

By using the $r$-matrix (\ref{3-para}), the AdS$_5$ part of \eqref{action} can be rewritten as 
\begin{align}
&L=L_G+L_B=
-\frac{1}{2} (\ga^{\al\be}-\ep^{\al\be}){\rm Tr}\left[A_\al P_2(J_\be)\right]\,, 
\label{action-abe3} \\ 
\text{with} \quad & 
J_\be\equiv \frac{1}{1-2\eta\bigl[R_{\rm Abe}^{(\mu_1,\mu_2,\mu_3)}\bigr]_g\circ P_2}A_\be\,,  
\end{align}
where $A_\al=g^{-1}\partial_\al g$ is restricted to $\alg{su}(2,2)$ and 
the R-operator associated with \eqref{3-para} is determined by the rule 
\eqref{linearR}\,.   

\medskip 

To evaluate the Lagrangian \eqref{action-abe3}\,, 
let us adopt the following coset parametrization \cite{ABF}\,:   
\begin{eqnarray}
g=\Lambda(\psi_1,\psi_2,\psi_3)\, \Xi(\ze)\, \check g_{\rm \rho}(\rho)    
\quad \in~ SU(2,2)/SO(1,4)\,. 
\label{coset-para} 
\end{eqnarray}
Here the matrices $\Lambda$, $\Xi$ and $\check g_\rho$ are defined as 
\begin{eqnarray}
&& \Lambda(\psi_1,\psi_2,\psi_3) \equiv \exp\left[\frac{i}{2}(\psi_1\,h_1+\psi_2\,h_2+\psi_3\,h_3)\right]\,, \el\\
&& \hspace*{-1cm}
\Xi(\ze) \equiv 
\begin{pmatrix} 
\cos\frac{\ze}{2} & \sin\frac{\ze}{2}& 0& 0 \\ 
-\sin\frac{\ze}{2} & \cos\frac{\ze}{2}& 0& 0 \\ 
0& 0&\cos\frac{\ze}{2} & -\sin\frac{\ze}{2}  \\ 
0& 0&\sin\frac{\ze}{2} & \cos\frac{\ze}{2} 
\end{pmatrix}\,, 
\quad 
\check g_\rho(\rho) 
\equiv\begin{pmatrix} 
\cosh\frac{\rho}{2} & 0& 0& \sinh\frac{\rho}{2}  \\ 
0 & \cosh\frac{\rho}{2} & -\sinh\frac{\rho}{2} & 0 \\ 
0& -\sinh\frac{\rho}{2} &\cosh\frac{\rho}{2}  & 0  \\ 
\sinh\frac{\rho}{2} & 0&0 & \cosh\frac{\rho}{2}  
\end{pmatrix}\,. \nonumber      
\end{eqnarray}

\medskip 

To find the projected current $P_2(J_\al)$\,, it is necessary to solve the following equation,  
\begin{align}
\left(1-2\eta P_2\circ \left[R_{\rm Abe}^{(\mu_1,\mu_2,\mu_3)}\right]_g\right)P_2(J_\al)=P_2(A_\al)\,. 
\label{equation2}
\end{align}
Note that $P_2(A_\alpha)$ is expanded with respect to the $\gamma$ matrices, 
\begin{align}
P_2(A_\al)&=\frac{1}{2}\Bigl[
-\partial_\al \rho \,\ga_1
+ i \cosh\rho\, \partial_\al\psi_3\, \ga_5\el\\
&\qquad 
-\sinh\rho( \cos\ze\,  \partial_\al\psi_1 \,\ga_2
+  \partial_\al\ze \,\ga_3
- i  \sin\ze\, \partial_\al \psi_2 \, \ga_0) \Bigr]\,. 
\label{P_2(A)2} 
\end{align}
Then, by combining (\ref{P_2(A)2}) with \eqref{equation2}\,, 
$P_2(J_\alpha)$ can be obtained as 
\begin{align}
P_2(J_\alpha)=j_\alpha^0\,\ga_0+j_\alpha^1\,\ga_1+j_\alpha^2\,\ga_2 
+j_\alpha^3\,\ga_3 +j_\alpha^5\,\ga_5\,, 
\end{align}
with the coefficients 
\begin{align}
&j_\alpha^0=-\frac{i}{2}
\frac{ \sin\ze \sinh\rho}{1 - 16 \eta^2[(\mu_1^2 \sin^2\ze+\mu_2^2 \cos^2\ze) \sinh^22\rho 
- \mu_3^2 \sin^22\ze \sinh^4\rho]} \el \\
&\qquad \times \Bigl[
(-1 + 16 \eta^2 \mu_2^2\cos^2\ze \sinh^22\rho)\partial_\al\psi_2 
\el \\
&\qquad \quad 
-8\eta (\mu_1- 8\eta\mu_2\mu_3\cos^2\ze \sinh^2\rho) \cosh^2\rho \partial_\al\psi_3
\el \\
&\qquad \quad 
-8\eta (\mu_3 - 8\eta\mu_1\mu_2 \cosh^2\rho) \cos^2\ze\sinh^2\rho\partial_\al\psi_1\Bigr]\,,  
\el \\ 
&j_\alpha^1=-\frac{1}{2}\partial_\al \rho\,, 
\el \\
&j_\alpha^2=\frac{1}{2}
\frac{\cos\ze\sinh\rho}{1 - 16 \eta^2[(\mu_1^2 \sin^2\ze+\mu_2^2 \cos^2\ze) \sinh^22\rho 
- \mu_3^2 \sin^22\ze \sinh^4\rho]} \el \\
&\qquad \times \Bigl[
(-1 + 16 \eta^2 \mu_1^2 \sin^2\ze \sinh^22\rho) \partial_\al \psi_1  
\el \\
&\qquad \quad 
+8 \eta ( \mu_3 +8\eta\mu_1\mu_2\cosh^2\rho)\sin^2\ze \sinh^2\rho \partial_\al\psi_2
\el \\
&\qquad \quad 
+8\eta (\mu_2+8\eta\mu_1\mu_3\sin^2\ze \sinh^2\rho )\cosh^2\rho\partial_\al\psi_3 
\Bigr]
\,, \nonumber \\
&j_\alpha^3=-\frac{1}{2}\sinh\rho \partial_\al\ze\,, \el\\
&j_\alpha^5=\frac{i}{2} \frac{\cosh \rho}{1 - 16 \eta^2[(\mu_1^2 \sin^2\ze+\mu_2^2 \cos^2\ze) \sinh^22\rho 
- \mu_3^2 \sin^22\ze \sinh^4\rho]} 
\el\\
&\qquad \times \Bigl[
(1 + 16 \eta^2\mu_3^2 \sin^22\ze \sinh^4\rho)\partial_\al\psi_3 
\el \\
&\qquad \quad 
-8 \eta (\mu_2-8 \eta \mu_1\mu_3 \sin^2\ze \sinh^2\rho) \cos^2\ze\sinh^2\rho \partial_\al\psi_1
\el\\
&\qquad \quad 
+8\eta(\mu_1+ 8\eta\mu_2\mu_3 \cos^2\ze \sinh^2\rho) \sin^2\ze \sinh^2\rho \partial_\al\psi_2 
\Bigr]  \,. 
\end{align}
Finally, $L_G$ and $L_B$ are given by, respectively,   
\begin{align}
L_G &= -\frac{\ga^{\al\be}}{2} \Bigl[
-\sinh^2\rho \partial_\al\rho\partial_\be\rho  
\el\\
&\qquad\quad  
+(\sin\ze \sinh\rho \partial_\al\ze -\cos\ze \cosh\rho\partial_\al\rho)
(\sin\ze \sinh\rho \partial_\be\ze -\cos\ze \cosh\rho\partial_\be\rho)
\el\\
&\qquad\quad  
+(\cos\ze \sinh\rho \partial_\al\ze + \sin\ze\cosh\rho \partial_\al\rho )
(\cos\ze \sinh\rho \partial_\be\ze + \sin\ze\cosh\rho \partial_\be\rho )
\el \\
&\qquad\quad 
+\hat G\bigl[
 \sinh^2\rho (\cos^2\ze \partial_\al \psi_1\partial_\be \psi_1 
+\sin^2\ze \partial_\al\psi_2\partial_\be\psi_2) 
- \cosh^2\rho \partial_\al \psi_3\partial_\be \psi_3 
\el \\
&\qquad\qquad 
-\cos^2\ze\sin^2\ze  \cosh^2\rho \sinh^2\rho 
\bigl(\textstyle{\sum_{i}}\hat\ga_i \partial_\al \psi_i\bigr) 
\bigl(\textstyle{\sum_{j}}\hat\ga_j \partial_\be \psi_j\bigr)\bigr] 
\Bigr]\,,  
\label{LG3ads}
\\
L_B&=-\ep^{\al\be}\hat G  \Bigl[
\hat\ga_3 \cos^2\ze \sin^2\ze \sinh^4\rho \partial_\al\psi_1 \partial_\be\psi_2
\el\\
&\qquad\qquad
-\sinh^2\rho \cosh^2\rho \left(\hat\ga_2 \cos^2\ze \partial_\al\psi_3 \partial_\be\psi_1
+ \hat\ga_1 \sin^2\ze \partial_\al\psi_2 \partial_\be\psi_3 \right)\Bigr]\,. 
\label{LB3ads}
\end{align}
Here a scalar function $\hat G$ is defined as   
\begin{align}
\hat G^{-1}\equiv 1 - (\hat\ga_1^2\sin^2\ze+\hat\ga_2^2 \cos^2\ze ) \cosh^2\rho \sinh^2\rho  
+ \hat\ga_3^2\cos^2\ze \sin^2\ze \sinh^4\rho\,,  
\end{align}
and new deformation parameters $\hat\ga_i$ are  
\begin{align} 
\hat\ga_i\equiv 8\eta\,\mu_i   \qquad (i=1,2,3)\,. 
\end{align}

\medskip 

By performing the coordinate transformation \eqref{coord-trf}\,,    
the metric and NS-NS two-form associated with \eqref{LG3ads}
and \eqref{LB3ads} are written into a compact forms,  
\begin{eqnarray}
ds^2 &=& \sum_{i=1}^3 (d\rho_i^2+\hat G \rho_i^2d\psi_i^2) 
+\hat G \rho_1^2\rho_2^2\rho_3^2 \left(\sum_{i=1}^3\hat{\gamma}_i\, d\psi_i\right)^2
+ds^2_{\rm S_5}\,,   
\label{3-metric} \\ 
B_2 &=& \hat G \left(
\hat{\gamma}_3\,\rho_1^2\rho_2^2\,d\psi_1\wedge d\psi_2 
+ \hat{\gamma}_1\,\rho_2^2\rho_3^2\,d\psi_2\wedge d\psi_3  
+ \hat{\gamma}_2\,\rho_3^2\rho_1^2\,d\psi_3\wedge d\psi_1 
\right)\,.
\label{3-NSNS}
\end{eqnarray}
Here there is a constraint $\sum_{i=1}^3 \rho_i^2=-1$ 
and $\hat G$ turns out to be   
\begin{eqnarray}
\hat G^{-1} = 1 + \hat{\gamma}_3^2\,\rho_1^2\rho_2^2 
+ \hat{\gamma}_1^2\,\rho_2^2\rho_3^2 
+ \hat{\gamma}_2^2\,\rho_3^2\rho_1^2\,. 
\end{eqnarray}
These are quite similar to $\gamma$-deformed S$^5$ \cite{Frolov,LM} 
and hence the metric (\ref{3-metric}) and NS-NS two-form \eqref{3-NSNS} may be 
regarded as $\ga$-deformed AdS$_5$\,.  

\medskip 

The one-parameter result in section 4 is reproduced by setting the parameters as 
\begin{align}
\hat\ga_1=\hat\ga_2=0\,, \qquad \hat\ga_3 =\hat\ga\,. 
\end{align}

\end{document}